# Neuronal micro-culture engineering by microchannel devices of cellular scale dimensions

*Gaurav Goyal and Yoonkey Nam*

*Department of Bio and Brain Engineering, KAIST, Daejeon 305-701, South Korea*

**Abstract**

Purpose
The purpose of the current study was to investigate the effect of microchannel geometry on neuronal cultures and to maintain these cultures for long period of time (over several weeks) inside the closed microchannels of cellular scale dimensions.

Methods
The primary hippocampal neurons from E-18 rat were cultured inside the closed polydimethylsiloxane (PDMS) microchannels of varying sizes. The effect of the channel geometry on the spatial and the temporal variations in the neural microenvironment was investigated by studying neural maturation and variation in the media osmolality respectively. The cultures were maintained for longer time spans by PDMS device pretreatment, control on media evaporation (by using hydrophobic ethylene propylene membrane) and an effective culture maintenance protocol. Further, the devices were integrated with the planar microelectrode arrays (MEA) to record spontaneous electrical activity.

Results
A direct influence of channel geometry on neuron maturation was observed with cells in smaller channels maturing faster. The temporal variation in the microenvironment was caused by several fold increase in osmolality within 2-3 days due to rapid media evaporation. With our culture methodology, neurons were maintained in the closed channels as small as 50 μm in height and width for over 1 month in serum free media condition and the time varying spontaneous electrical activity was measured for up to 5 weeks using the MEA.

Conclusions
The understanding of the effect of the culture scale on cellular microenvironment and such long-term culture maintenance will be helpful in studying neuronal tissue development; therapeutic drug screening; and for network level neuronal analysis.

**Introduction**

The usefulness and reliability of the *in vitro* cell culture models depend on their ability to capture the characteristic tissue physiology in terms of cell-cell interaction, cell extracellular matrix interaction and cellular chemical and mechanical microenvironment. However, the conventional cell culture models currently in use are limited in their ability to mimic the *in vivo* cellular microenvironment owing to the large culture scale relative to the cell size, the low cell volume to the extracellular fluid volume ratio and the convective media flow [1-3]. The escalating understanding of cellular microenvironment *in vitro* and the advancements in microtechnology have led to the development of microscale cell culture models over the past few years [4, 5]. The miniaturized architecture of these microdevices renders the cell culture scale to be proportional to the spatial resolution of cell-to-cell communication and cell volume to the extracellular fluid volume ratio greater than one [1], thereby establishing sustained chemical gradients and simulating the cellular microenvironment analogous to *in vivo*.

Such miniaturized cell culture devices can also be used to grow nerve cells in confined spaces with tens of nano-liters of culture media. There have been some attempts to culture mammalian neurons in closed polydimethylsiloxane (PDMS) microfluidic channels for various neurobiological applications. Taylor et al. have reported a microfluidic neuron culture device composed of two large fluidic compartments (width: 1.5 mm, height: 100 µm, length: 8 mm) connected by microchannels. They selectively guided axons from one compartment to the other and used the device for axonal injury and regeneration studies [6, 7]. The similar culture device has also been used by some other groups for studying injury and regeneration of individual axons [8], culturing neurons derived from embryonic stem cells [9], for purification of the axonal mRNA [10], for electrophysiological measurements [11], and to study intracellular pH regulation in the neurites and the soma [12]. Recently, low-density neuronal cultures inside the narrow PDMS microchannels (channel width: 85 µm, height: 45 µm, length: variable) have also been reported [13]. Millet et al. proposed a series of chemical treatment of the PDMS to enhance neuronal survival in the close-channel devices; however, their neuronal cultures inside the closed microchannels could be maintained only up to 7 days in the static bath devices and up to 11 days in the perfusion devices made from native, autoclaved or the chemically treated PDMS.

All of the above discussed investigations have presented a new approach for neuronal analysis *in vitro*; however, there is only a limited knowledge available about the effect of microchannel geometry on shaping up the cellular microenvironment, and its influence on the neuronal cultures inside the channels. Furthermore, the reported works have not been very successful in maintaining neurons in the narrow microchannels for long period of time, thereby precluding their use for longitudinal neuronal studies *in vitro*. It is difficult to understand the complex cellular microenvironment simulated inside the microchannels because of its inaccessibility; nevertheless, this knowledge is vital in order to design the effective and reliable microscale neuron culture models and to engineer the neural microenvironment *in vitro*. The other important prerequisite for the usefulness of such culture models is the ability to culture neurons for a 'long time' so that these models can also be used for a variety of *in vitro* neural analysis which is currently possible only with the conventional culture systems (using culture flasks and Petri dishes).

Here we report the investigation of micro-scale neuronal cultures engineered in microchannels of cellular dimensions. PDMS microchannel devices were fabricated by soft-lithography and E18 rat hippocampal neurons were cultured up to one month in close channel devices. Early neuronal development and long-term electrophysiological properties were characterized. The size of the culture channels was found to be significant parameter and the narrower channels resulted in the faster neuron maturation. By adequate pretreatment of the PDMS devices, efficient control over media evaporation and an effective culture maintenance protocol, we maintained the micro-cultures for over one month in as small as 50 µm high, 50 µm wide and 7 mm long closed microchannels. Further, we successfully integrated the microchannel cultures with multi electrode arrays (MEAs) and recorded the spontaneous extracellular activity over a span of 5 weeks. Figure 1 shows the experimental scheme for this work.

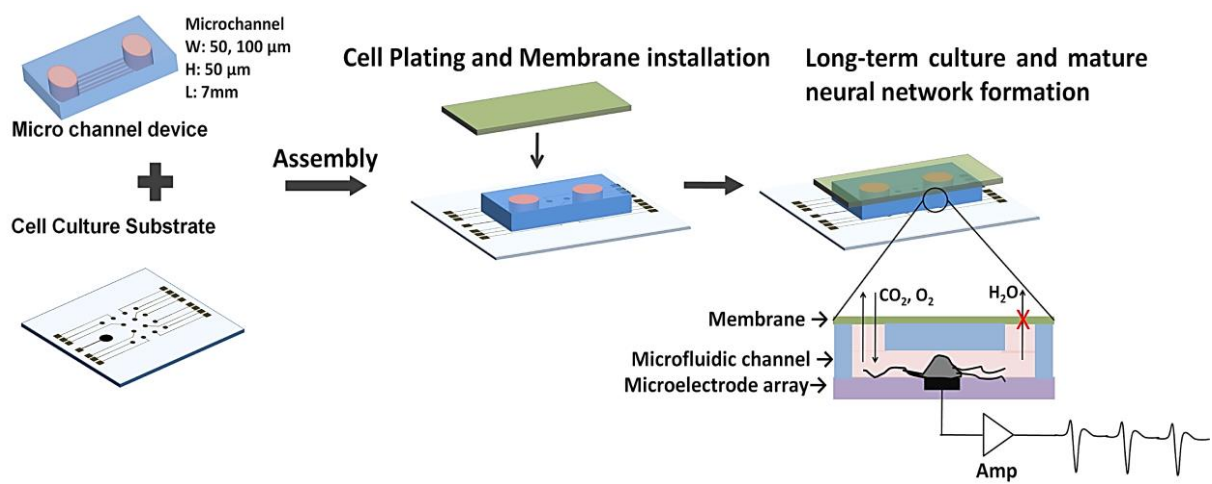

*Figure 1* Schematic for microdevice integration with microelectrode arrays. Microchannel devices were assembled on poly-D-lysine coated substrates followed by cell plating in the device and membrane installation. After the network maturation, MEA could be used to record signals from neurons inside the microchannels.

**Methods**

**Microchannel device fabrication, preparation and assembly**

The masters for the microchannel devices were fabricated by two methods and were divided into two groups (Table 1).

**Table 1** Microchannel dimensions used for maturation study.

|         | Width (μm) | Height (μm) | Length (mm) |
|---------|------------|-------------|-------------|
| Group 1 | 1000       | 150         | 7           |
|         | 1000       | 50          | 7           |
| Group 2 | 200        | 50          | 7           |
|         | 100        | 50          | 7           |
|         | 50         | 50          | 7           |

The group 1 consisted of 1 mm wide single channel devices fabricated by a clean-room free process as illustrated in Figure 2. The channel features were made by taping 3M Scotch transparent tape (MMM600341296, 3M, St. Paul, MN) on clean glass slides. A single layer of tape resulted in 50 μm high features and multiple layers were used to fabricate higher features. The second group consisted of the devices with 10 parallel microchannels spaced by 200 μm. For each channel, the height was 50 μm and the width was 50, 100 or 200 μm. The masters were fabricated using standard photolithography process. Briefly, a 4-inch silicon wafer was coated with the negative photoresist SU-8 2025 (MicroChem Corp., Newton, MA), soft baked and exposed through a film mask using Q4000 Infrared Mask Aligner (Quintel Corp., Morgan Hill, CA). A positive relief pattern corresponding to the microchannel geometry was obtained after the post-exposure bake and development.

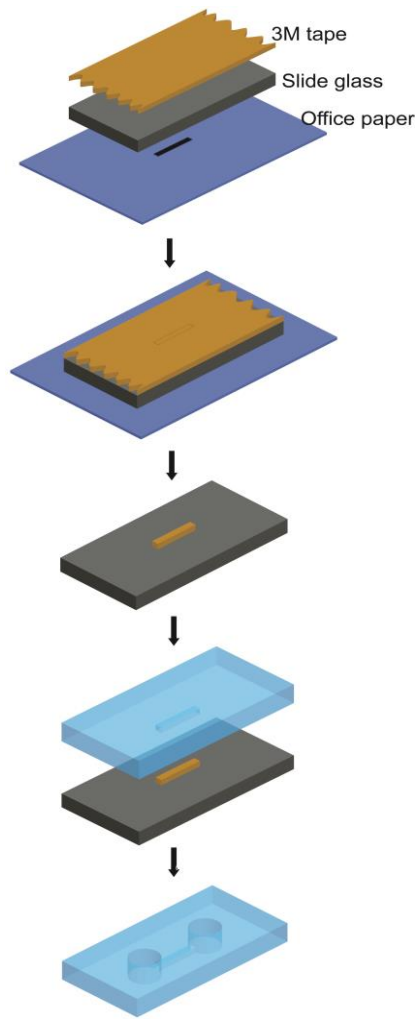

***Figure 2*** *Clean-room free fabrication process. 3M Scotch transparent tape was pasted on a cleaned glass slide and was assembled on a normal office paper with straight channel pattern printed on it. The tape on the glass slide was cut using a surgical knife following the channel pattern visible under the glass slide. The background tape was removed to obtain positive relief pattern on glass. PDMS was cast against the tape master and holes were punched in the molded PDMS to obtain the desired device.*

After the masters were fabricated, PDMS (Sylgard 184, Dow Corning Corp., MI) was cast against the masters to create desired microchannel devices. Input and output ports were made by punching holes (3 mm or 7 mm diameter) in the fabricated PDMS devices. The devices were then dipped overnight in methanol to remove any uncured PDMS, followed by thorough washing in deionized water (4-5 hours dipping) and autoclaving. The inner walls of the PDMS channels were made selective hydrophilic by assembling the devices on a plain blank glass with the feature side down and treating with air plasma (Harrick Expanded plasma cleaner, 3 min at 30 W, ~200 mTorr). The devices were then peeled off from the glass and were reversibly bonded on the poly-D-lysine (100 μg/ml, Sigma-Aldrich, St. Louis, MO, US) coated cell culture substrates (glass cover-slip or microelectrode array). The channels were immediately filled with the cell plating media and the assemblies were housed in the Petri dishes and were placed in a humidified incubator until cell plating.

**Cell Culture**

For cell culture, E-18 hippocampal neurons from Sprague-Dawley (SD) rat (Koatech Animal Services, Pyeongtaek-si, Gyeonggi-do, Korea) were suspended in cell plating media consisting of Neurobasal media (Invitrogen, Gaithersburg, MD, US) supplemented with B-27 (Invitrogen), 2 mM GlutaMAX (Invitrogen), 12.5 μM L-glutamic acid (Sigma-Aldrich, St. Louis, MO, US) and Penicillin-Streptomycin (1:100 , Invitrogen). For cell loading, the devices kept in the incubator were taken out and the excess media from the reservoirs was aspirated. The cell suspension of $3 \times 10^6$ cells/ml was prepared and was loaded through one of the wells to achieve a cell density of 250-300 cells/mm$^2$ inside the channels. High density cultures were prepared for the electrophysiological experiments to ensure signal recording. The devices were then again kept inside the humidified incubator maintained at 37°C and 5% $CO_2$ level. Additional culture media was added to the devices after 20 minutes of cell settling.

**Maturation assessment**

During early maturation of the neurons *in vitro*, 5 different stages have been reported [14]. Over the period of one week after cell plating, neurons undergo several morphological changes before they attain the functionally mature state and become electrically active. We observed that all the cells do not grow at the same pace but they are distributed from maturation stage 1 to stage 3 at 1 day *in vitro*. To explore the effect of culture vessel geometry on neuron maturation, we calculated the percentage of neurons in different maturation stages growing inside channels of different dimensions. To calculate the maturation index, the bright-field phase contrast images of the neurons were taken at 1 day *in vitro* (DIV) and the cells were classified into different stages following their morphological features. At stage 1, the soma was visible with lamellipodial projections and no neuritees; at stage 2, the minor processes projected from the soma; stage 3 was marked by one major neurite which would later become the axon of the cell. The cells were considered in stage 3 if their major neurite was more than twice as long as the other neurites. The classification of neurons was performed by an individual blinded to the experimental design to avoid the bias in data collection. The images of the neurons were taken across the complete length of the channel to avoid any bias originating from location of the neurons inside the channels. Only the neurons which could easily be classified in the 3 maturation stages were considered and any cell clusters were neglected from the analysis.

**Cell Culture Maintenance**

The rapid evaporation of media from the devices is a significant problem for microscale cell culture. We hypothesized that the continuous loss of aqueous phase from the devices was a major perturbation in cellular microenvironment in our cultures. To investigate this temporal variation in microenvironment due to media evaporation, the osmolality of the media in the devices was regularly measured using the freezing point depression osmometer (K-7400, Knauer, Germany). Further, to address the problem of evaporation, the devices were installed with a hydrophobic fluorinated ethylene propylene (FEP) membrane [15] (ALA Scientific Instruments, New York, US) after cell plating. The FEP membrane was introduced by Potter et al. to prevent media evaporation from their cultures with 1 ml media and is generally used in the labs working with MEA. We expanded the use of this hydrophobic membrane to microdevices in order to minimize the loss of culture media, thereby maintaining a favorable chemical microenvironment for healthy cultures. For the membrane installation, FEP membrane was carefully placed on the PDMS devices and was gently pressed onto the PDMS surface with a tweezers. The membrane could easily stick to the hydrophobic PDMS surface and make a seal which prevented the media from evaporation. The membrane could be easily peeled off and reinstalled while changing the media in the devices. For the culture nourishment, media in the devices was replaced with the maintenance media (Neurobasal media supplemented with B27, Glutamax, and Penicillin-Streptomycin) after 2 days of cell plating and then half of the media was changed every 3-4 days to replenish the depleted nutrients and to remove the accumulated wastes [16, 17].

**Electrophysiology**

To measure electrical activity from the neurons growing inside the channels, the PDMS microchannel devices were integrated with the planar microelectrode array (Multichannel systems, Reutlingen, Germany) and MEA 1060 System (Gain 1200, Bandwidth: 10 – 5000 Hz, Multichannel Systems) was used to collect multichannel extracellular neural signals ('spikes'). The spike detection threshold was 5 standard deviations of the background noise and well isolated unit spikes were sorted out by Offline sorter (Plexon Inc., Dallas, TX, USA). The spike frequency and the burst frequency were calculated using NeuroExplorer (Nex Technologies, Littleton, MA) following the surprise algorithm for the burst detection (minimum surprise = 2).

## Results

**Effect of microchannel geometry on neuron maturation**

In order to explore the effect of microchannel geometry on maturation of neurons, the cells were grown in conventional Petri dish (35 mm diameter with 2 ml culture media) and group 1 & 2 microchannel devices. Figure 3a shows the 5 maturation stages of neurons in the culture as described by Dotti et al [14]. These stages were used to classify the neuron population in our devices. Figure 3b shows the neurons at 1DIV distributed into stages 1-3. The neurons in Figure 3b are marked as S1, S2 and S3 for Stage 1, Stage 2 and Stage 3 respectively. The maturation index was calculated for all of our culture devices and is shown in the Figure 3c.

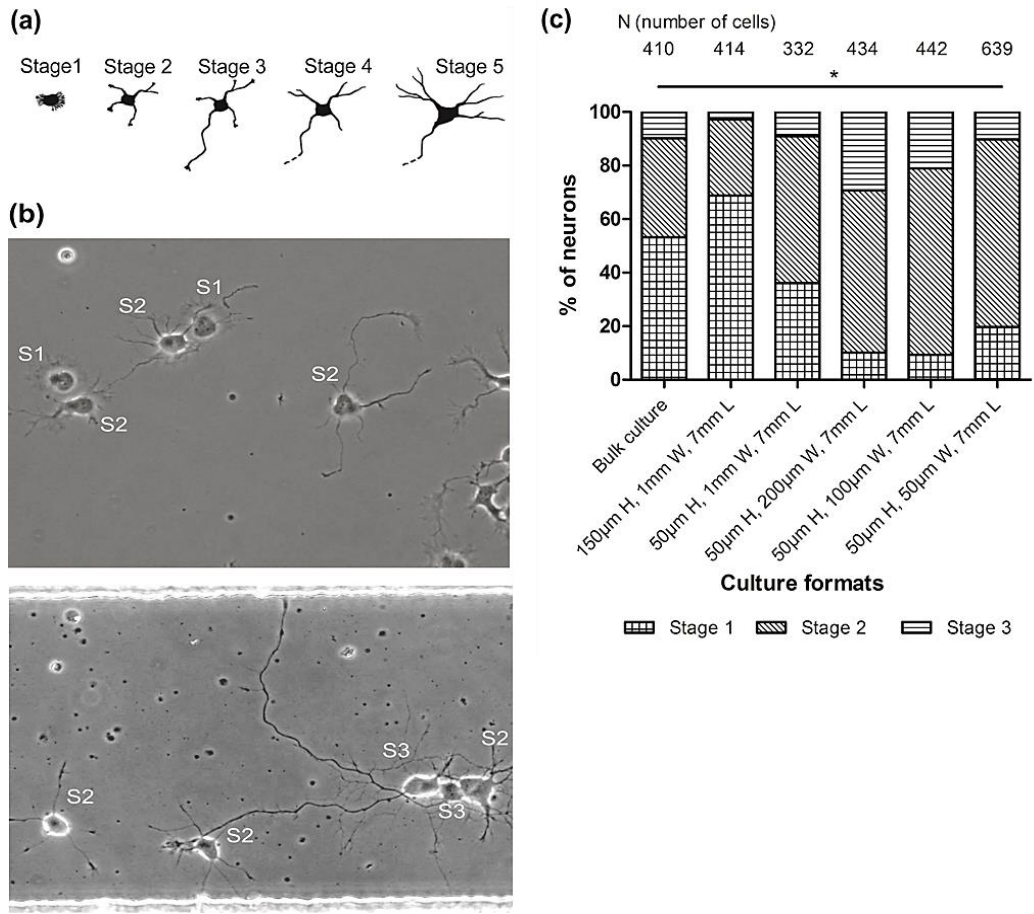

***Figure 3*** *Maturation of neurons in microchannels. (a) 5 different developmental stages of neurons in culture, used to classify the neuron population. The image was redrawn from Ref 14. (b) Images showing neural population distributed from stage 1-2 at 1DIV. The neurons in stage 1, 2 and 3 are marked as S1, S2 and S3 respectively. (c) Distribution of cells in different maturation stages in different culture formats. The chi-square statistic was calculated for the data and there is significant evidence with \*p<0.01 that the distribution of cells in different stages is different in all the culture formats.*

For the conventional Petri dish, out of 410 cells analyzed 53.2 % neurons were classified in stage 1 and only 37.1 % neurons were classified in stage 2 and rest in the stage 3. For the microdevices with 1 mm wide and 150 μm high channels, we observed a slight increase in the stage 1 population as compared to Petri dish culture. We observed 68.8 % cells in stage 1 and 28.5 % cells in stage 2 and a very few cells in stage 3 out of the total 414 cells analyzed for these devices. But when the dimensions of the channels were brought further down, the stage 2 neurons became most predominant in the culture. It suggested the role of miniaturized dimensions of the culture vessels in accelerated neuron maturation. For the devices with 1 mm wide and 50 μm high channels, out of 332 neurons 36.1 % were classified in stage 1, 54.8 % in stage 2 and the rest in the stage 3. This switch from stage 1 to stage 2 was seen as an effect of lowering the channel height with the constant channel width. When the width of the channels was also reduced to 200 and 100 µm, there was further rise in stage-2 population and a rise in stage-3 population of the neurons. For the devices with 200 µm wide and 50 µm high channels, out of 434 neurons only 10.1 % cells were in stage 1, and 60.6% and 29.3 % neurons were in stage 2 and 3 respectively. The devices with 100 µm wide and 50 µm high channels had 9.3 % cells in stage 1, 69.7 % cells in stage 2 and 21.0 % neurons in stage 3 of the total 442 cells analyzed. This trend of accelerated maturation upon reducing the channel dimensions was not maintained when the width of the channels was reduced to 50 µm and we observed increase in stage 1 population compared to the devices with 100 µm width. Out of 639 neurons inside channels with 50 µm height and width, 19.7 % were classified in stage 1, 70. % in stage 2 and the rest in stage 3. These results suggested that the height of microchannels was a critical design parameter which significantly altered the spatial chemical microenvironment of the neurons. This observation is in agreement with previous report by Yu et al. [2] in which they indicated the height of microchannel as a governing parameter in diffusion based microenvironment simulated in the channels. Our results also suggested that the optimum width of the channels should be in the range of 100 - 200 µm for maintaining a favorable microenvironment for neuron growth. We calculated chi-square statistic for all the culture formats and it reflected strong statistical evidence with p-value <0.001 that the 6 culture devices were different. We also did a series of follow up chi-square tests by taking one of the devices out at a time from the analysis to see if the variability was due to any one device. The results suggested that all the culture devices were different and variability was not due to any one of them.

**Osmolality maintenance and neuron survival**

The designed devices consisted of microchannels of cellular scale dimensions (down to 50 µm height and width) which significantly reduced the media required for the culture. The total amount of media contained in the devices was 50 or 200 µl (corresponding to 3 mm or 7 mm wide media reservoirs); and routinely the 50 µl media evaporated in a single day from the devices kept inside the humidified incubator which resulted in early cell death. The devices with 200 μl media endured complete evaporation for 3 - 4 days and in these devices media was replenished before drying out, but the neurons still could not survive for over 2 weeks of duration (data not shown).

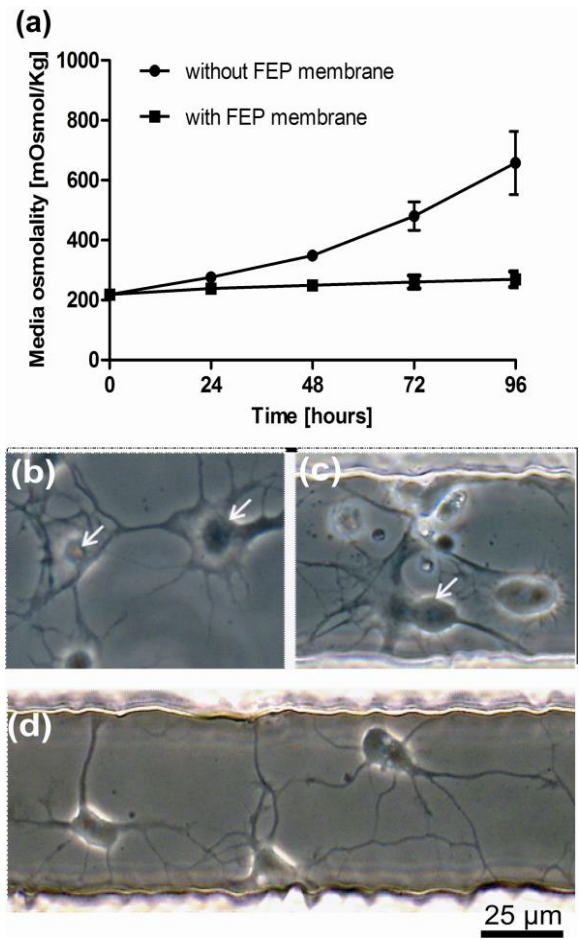

*Figure 4 Variation of media osmolality in microdevices and its effect on cell health. Media evaporation from microdevices caused the osmolality to rise to three fold of the initial value within 4 days but after membrane installation osmolality was well maintained (a). In hyperosmolar media conditions cells appeared flattened and the cell nucleus condensed and became distinctively visible (marked by white arrows) by 2 DIV. (b) & (c) show degrading cells in 100 and 50 μm wide channels respectively. After membrane installation, media osmolality was maintained and cells appeared round and healthy (d). Cells shown are at 2 days in vitro.*

In order to prevent the loss of aqueous phase from the culture media, the devices were installed with a water impermeable FEP membrane. After the membrane installation, the original media volume was maintained in the devices and it helped to preserve the osmolality of the media. Figure 4 shows the variation in media osmolality in the microchannel devices with and without the membrane and its effect on the morphology of the growing neurons. Without the membrane, the osmolality increased to 3 folds (657.5±74.5 mOsmol/kg) of the initial value (219±1 mOsmol/kg) in 4 days. Such a large change in osmolality had significant effects on the health of the neurons inside the microchannels; the neurons became more flattened, lost their membrane integrity and the cell nucleus shrank and became conspicuously visible by 2 days after the cell plating (Figure 4b and 4c). Such cell morphology can be attributed to the loss of cytosol by the neurons due to the hyperosmolar media condition [18-20]. When the FEP membrane was installed, the osmolality could be successfully maintained within physiological range (269.5±25 mOsmol/kg) even after 4 days. In the devices with FEP membrane, neurons exhibited healthy morphology and maintained the spherical cell shape (Figure 4d) and they could be maintained for several weeks in the microchannels. These observations implied that merely changing the media frequently was not sufficient to sustain the neural culture in healthy state; rather, an effective means to prevent media loss was needed to maintain the favorable physiological conditions.

**Neuronal characteristics and long-term culture**

The cells were labeled with the antibodies against the neuronal marker class III β-tubulin to observe neuron development and the positive staining was obtained (Figure 5a, cells in a 50 µm wide channel at 3DIV) at different time points confirming that the cells normally developed to express the neuronal markers and maintained their neuronal characteristics (data not shown). With the control on cellular microenvironment, the low density neuronal cultures could be stably and reproducibly maintained for several weeks inside the closed channels. Figures 5b & 5c show single neurons inside a 50 µm wide (and 50 µm high) channel at 6 DIV and 28 DIV respectively. Figure 5d shows a 6 DIV old neuron in a 100 µm wide (50 µm high) channel. Figure 5e is the representative image for our long-term culture, showing a 33 DIV old neuron in a 100 µm wide and 50 µm high channel. The characteristic morphology of hippocampal neuron was observed, marked by triangular shaped neurons, along with regular spread and branching of neurites. The neurites were confined to the microchannels and did not grow under the PDMS constraints.

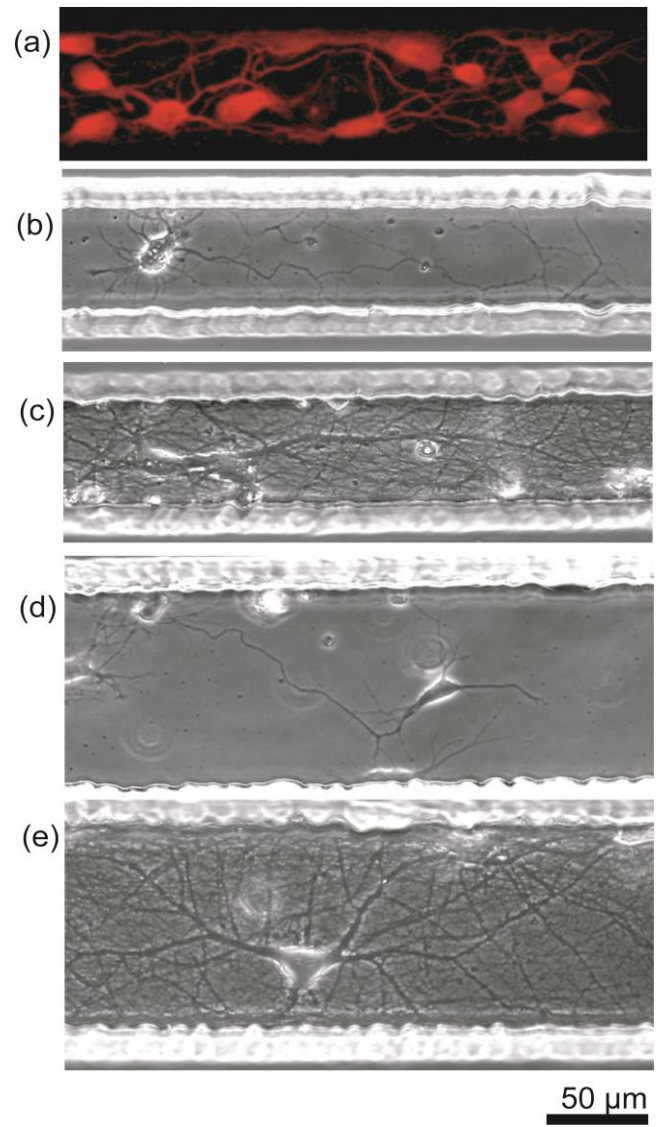

*Figure 5* *Neuron development in microchannels and long term culture. Neurons in microchannels showed characteristic growth of hippocampal pyramidal neurons. (a) Neurons in 50 µm wide channels stained for neuronal marker class III β- tubulin at 3 DIV. (b) & (c) Single neurons inside 50 µm wide channels at 6 and 28 DIV respectively. (d) & (e) 6 DIV and 33 DIV old neurons inside the 100 µm wide channels.*

**Multichannel neural recordings for the development of neural networks**

In order to demonstrate the feasibility of long-term network-level physiology experiments with the present technique, we integrated our microchannel devices with the planar multi-electrode arrays that are commonly used for multichannel neural recordings of *in vitro* neural tissues. For coupling with the MEA, the microchannel devices were manually aligned with MEA electrodes using an inverted microscope and were bonded reversibly. A total of 8 devices (with 50 or 100 µm wide channels) were coupled with the MEA. Out of the total 8 microdevices, spontaneous activity from 4 devices could be recorded up to 5 weeks *in vitro*. The activity from other 2 devices was recorded for 3-4 weeks and the rest 2 devices did not show any activity. Figure 6 presents the representative images and recording data from our electrophysiology experiment. Figure 6a shows the confined neural networks growing inside 8 different microchannels (100 μm in width) at 17 DIV. Figure 6b shows a 100 μm-wide microchannel aligned to 30 μm-TiN electrodes and dense neurites or somata growing over the electrodes owing to the confined neural adhesion and growth. Figure 6c represents multichannel neural recordings from three neighboring surface-embedded microelectrodes. The typical noise level estimated from the electrodes was 18 – 20 $\mu V_{rms}$ and the detected spikes were mainly 'negative' spikes with their peak values ranging from – 100 μV to – 700 μV. For example, the biggest spikes in Figure 6c (marked with the dots) had negative peak values of – 374.2 μV (7 DIV), – 561.7 μV (17 DIV), and – 682.5 μV (28 DIV). Although the measured noise level was much higher than the level without channels (2 – 3 $\mu V_{rms}$), relatively large spikes resulted in the high signal-to-noise ratio for reliable spike detection.

The capability of long-term culture allowed us to observe the maturation of the network activity through the electrophysiological recordings. As the confined neuronal networks matured, the number of spikes generated from the individual neurons ('spike rates') and the degree of temporal clustering of the spikes ('burst') also increased. The estimated values are summarized in table 2. There was a several fold increase in the network activity from 7 DIV to 28 DIV. The increase in the network activity after 2 weeks is a typical characteristic of the cultured neural networks and can be attributed to synapse maturation and increase in the synapse density with time [21]. These results demonstrate the possibility to perform the same level of electrophysiological experiments with neuronal cultures inside closed microchannel devices as it is possible with macro-scale cultures.

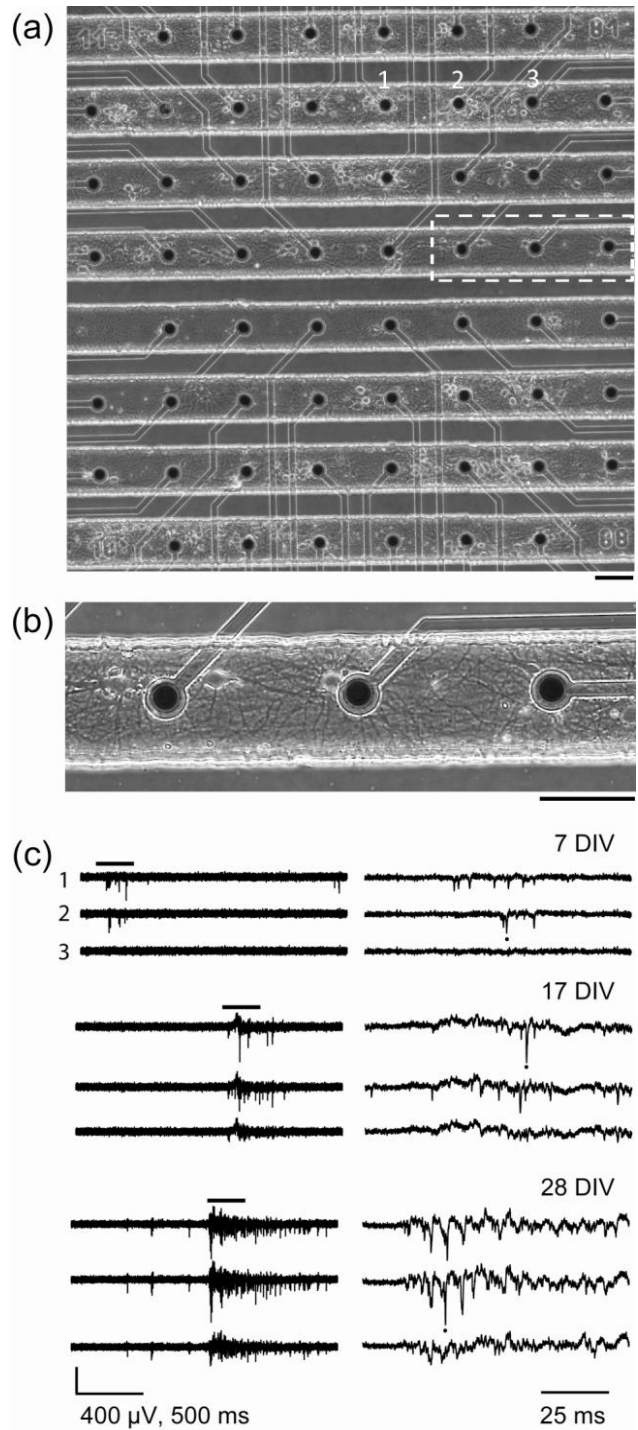

***Figure 6*** *Microchannel culture integration with MEAs. (a) The MEA electrodes overlaid with a 100 μm wide channels and cultured neural networks at 17 DIV. Scale bar 100 μm. (b) Magnified view of three neighboring electrodes inside the channel marked with dashed lines in (a). Scale bar 100 μm. (c) Electrical activity recorded from the three electrodes (marked as 1, 2 and 3 in part (a)) at 17, 17, and 28 DIV. Left: 2 sec traces. Right: zoomed traces (100 ms) of the marked segments in left traces.*

**Table 2** Estimated spike rates and bursts rates in Figure 6c at three different culture ages

| Electrode | Spike rate [spikes/sec] | | | Burst rate [bursts/min] | | |
|---|---|---|---|---|---|---|
| | 7 DIV | 17 DIV | 28 DIV | 7 DIV | 17 DIV | 28 DIV |
| 1 | 0.355 | 1.668 | 7.085 | 2.16 | 4.56 | 17.46 |
| 2 | 0.127 | 1.171 | 8.638 | 0.96 | 3.90 | 23.22 |
| 3 | - | 0.085 | 2.573 | - | 0.42 | 8.04 |

**Discussion**

**Neuron maturation study in microchannels**

The maturation of neurons *in vitro* is influenced by several factors, out of which cell-cell contact and the presence of neurotrophic factors are the major contributing elements. Neurons with a high concentration of growth factors and extracellular matrix components around them exhibit accelerated maturation in culture [22]. The microchannel culture model has been proposed to maintain a high concentration of intrinsic growth factors around the cells and to preserve their cellular microenvironment as compared to the Petri dish culture in which the convective media flow and the large media volume dilute the trophic factors emanated from the cells and take them away from the cell's immediate vicinity [2, 3]. Our rationale was that the spatial distribution of the neurotrophic factors around the cells depends on the dimensions of the culture vessels. The narrower the culture channel, the more the neurotrophic factors accumulated around the cells. So we started with the 1 mm wide and 150 μm high channels and then kept reducing the channel dimensions to investigate the effect this variation in channel geometry may have on neuronal maturation. Our results reflected that the trend was not linear for all the channel dimensions. When bulk culture was compared to the cultures inside 1 mm wide and 150 μm high channels, there were more cells in stage 1 in the channel culture than in the bulk culture. This can be explained by the hypothesis that the concentration of the neurotrophic factors is not significantly high (inside these channels, when compared to the bulk culture) to accelerate the cell maturation; rather, the negative factors such as waste accumulation and depleted nutrient condition come into play to pull the maturation rate down. When the channel dimensions are further reduced, we observed increase in the proportion of cells in stage 2 and stage 3 with each subsequent reduction in size of the channels. This observation of accelerated growth is in agreement with our rationale of designing this study. We believe for these three channel sizes, the increasing concentration of the neurotrophic factors (with the decreasing size of the channels) outweighs the negative factors of microchannel culture mentioned above and this makes the neurons to mature faster in these channels These results suggest that the design the microscale culture vessels for neuronal culture should be based on the optimum dimensions of the channels which can help to emulate the *in vivo* condition but have the minimal technical limitations of microscale culture system (high concentration of uncured polymer, cellular wastes and depleted nutrients; all arise due to small media volume).

The method we have used to investigate the effect of channel geometry on cellular microenvironment served as a good semi quantitative measure. The advantage of this means of quantification of neuron growth was that it was quick (because the neurons only needed to be classified into different growth stages based on the morphological features) and required only bright-field images. But it is a subjective scoring technique and is susceptible to bias if the experiment in not properly designed. The group 1 devices used for maturation study were fabricated with an unconventional clean room free fabrication method. Although this crude method of making the microchannels with tape masters can suffer from the drawbacks like batch to batch variations, limitation of resolution and the design complexity, it is however quick and easy. It provides a rapid prototype for microchannel system (within limited size range) which can be easily adopted by any conventional biology lab to explore a new paradigm of experimentation without substantial time and cost investments. Although the usefulness of the devices in this size range depends highly on the experiment design. The fabrication by this method does not require expensive set up or highly skilled personal. Tape masters can generally be prepared in a span of few minutes and are stable enough for multiple PDMS moldings. We have personally used our tape masters more than 20 times to mold the micro channels.

**Temporal maintenance of cellular microenvironment**

The chemical microenvironment generated around the cells inside the microchannels does not remain the same but changes with time. This temporal variation can be caused by media evaporation from the culture devices [6], leaching of cytotoxins from PDMS [1,13], waste accumulation and depleted nutrient and oxygen supply inside the channels [6,7,13]. We addressed the problems of the cytotoxin leaching from the PDMS by pretreatment and autoclaving the PDMS devices before cell culture. The other significant issue we encountered was the evaporation of media from the devices which increased the media osmolality to several fold of the initial value in a span of 3-4 days. Because the amount of media contained in the micro-devices is very small, the rise in media osmolality due to evaporation is much higher as compared to large scale cultures. This media loss problem is generally addressed by frequently adding extra media in the devices (or frequently changing the media in the devices) which replenishes the lost media volume [6]. But we believe when the media in the microdevices becomes hyperosmolar, frequently adding more media is not the optimum solution and can severely affect the cell health. When the cells growing in the hyperosmolar media are exposed to the fresh media (when media is changed/replenished), they are subjected to an osmotic shock. This osmotic shock is experienced by the cells every time the media is changed and it exposes the cellular membrane to osmotic stress. Such a media change protocol worsens the already unhealthy state of the cells caused by the hyperosmolar cellular environment and expedites cell death. In our opinion, effective control of media evaporation from the devices is an essential measure to maintain the favorable microenvironment around the cells for an extended duration. We installed the hydrophobic FEP membrane on our devices to prevent the evaporation and maintain the media osmolality in the physiological range. This temporal control over the cellular microenvironment was essential to maintain healthy neural cultures in the microchannels for long period of time.

**Need for the long term culture**

The long-term neuronal culture using the conventional culture model has been widely used to explore a variety of topics in neuroscience including synaptogenesis [23], neural aging and death [24], synaptic plasticity [25], and neural information processing [26]. For example, the expressions of glycolytic enzyme GAPDH, synaptic proteins (synaptophysin, α- and β-synuclein) and the level of glutamate receptors in the cultured neurons were shown to reach to the maximum expression level from 15 to 30 days after cell plating [24]. Moreover, β-synucleins which are implicated in neurodegenerative diseases such as Parkinson's disease and Alzheimer's disease were shown to be differentially localized in dendrites and axons in the immature (3DIV) and mature (20 DIV) neurons, respectively [24]. It has also been shown that neural network *in vitro* matures with culture age and starts to exhibit electrical activity after 2 weeks of cell plating [21,27], indicating 2 weeks of culture age to be minimum time-point for any meaningful network level electrophysiology experiment. It was therefore important to develop a reproducible methodology for culturing mammalian neurons in the microchannels for long period of time so that the microscale culture model can also be used for the similar kind of *in vitro* experimentation. Such a culture platform can enable the study of neural maturation, neural aging and death as it can facilitate definitive and long-term analysis of the chemical and bio-molecular signatures of neurons in the microchannels. It can also provide temporally rich information for drug experiments revealing the long term effects of drug exposure to micro scale cultures.

**Microdevice integration with the MEA**

We coupled our microchannel devices with the planar microelectrode arrays (MEA), to demonstrate an integrated device that may be used for the long-term noninvasive monitoring of neural activity and may serve as a means to study the synergistic effect of chemical and electrical stimulation on neuronal networks in the microchannels. Efforts have also been made previously to integrate commercial or custom-made MEAs with the microchannel neuron cultures [28-31]. Recently, Dworak and Wheeler also reported an MEA platform with the PDMS micro-tunnels to selectively record signals from the isolated axons [32]. In most of these previous works neurons were cultured in open wells with an easy access to the bulk media and only the neurites were exposed to the controlled microenvironment. The development of an integrated device with the closed microchannels has been pursued by many groups because it is an attractive platform for high-throughput drug screening and to study neurobiology. The electrodes underlying the neural circuits in microchannels can record action potentials and can be used to investigate the functional changes in the neural activity in a controlled microenvironment. So far, the success in implementing such a hybrid device had been limited by reliable long-term neuronal cultures in the microchannels to obtain functionally mature networks. Our microchannel cultures protocol has enabled us to culture the mammalian neurons inside the "closed microchannels" and to be able to record electrophysiological signals from these neurons over the period of 5 weeks. Moreover, we have integrated our devices with the commercially available MEA (Multichannel systems, Reutlingen, Germany) which allows other researchers to easily replicate, use and extend our work. Many diverse microfluidic techniques can also be integrated with MEA for novel neuroscience or cell based biosensor applications.

**Conclusion**

We cultured primary mammalian neurons inside the closed microchannels and investigated the effect of variation in the cellular microenvironment on our neural cultures. By effectively controlling the changes in neural microenvironment over time, we maintained the cultures for over 1 month in the serum free media. The long-term culture further enabled us to record spontaneous electrical activity from neural cultures when the microdevices were coupled with the microelectrode arrays. We believe this understanding of the cellular microenvironment and the methodology to maintain healthy long-term neuronal cultures will empower researchers to design more effective and reliable cell culture models for *in vitro* neuronal analysis.

**Acknowledgements**

This work was supported by Korea Research Foundation Grant (KRF-2007-211-D00123) and the Brain Research Center of the 21st Century Frontier Research Program from Ministry of Education, Science and Technology in Korea. Authors also thank Chung Moon Soul Center for Bioinforrmation and Bioelectronics at KAIST for the financial support.